\documentclass[sigconf]{acmart}

\acmConference[MSR 2022]{MSR '22: Proceedings of the 19th International Conference on Mining Software Repositories}{May 23–24, 2022}{Pittsburgh, PA, USA}

\usepackage{hyperref}
\hypersetup{colorlinks,allcolors=black}

\begin{document}
\title{Which bugs are missed in code reviews: An empirical study on SmartSHARK dataset}







\author{Fatemeh Khoshnoud}
\authornote{Three first authors contributed equally to this research.}
\email{khoshnoud.fa@gmail.com}
\affiliation{%
  \institution{Department of CSE and IT, Shiraz University, Shiraz, Iran}
  \city{}
  \country{}
}

\author{Ali Rezaei Nasab}
\authornotemark[1]
\email{alirezaei@hafez.shirazu.ac.ir}
\affiliation{%
  \institution{Department of CSE and IT, Shiraz University, Shiraz, Iran}
  \city{}
  \country{}
}

\author{Zahra Toudeji}
\authornotemark[1]
\email{zahra.toudeji94@gmail.com}
\affiliation{%
  \institution{Department of CSE and IT, Shiraz University, Shiraz, Iran}
  \city{}
  \country{}
}

\author{Ashkan Sami}
\authornote{Corresponding Author}
\email{sami@shirazu.ac.ir}
\affiliation{%
  \institution{Department of CSE and IT, Shiraz University, Shiraz, Iran}
  \city{}
  \country{}
}

\makeatletter
\let\@authorsaddresses \@empty
\makeatother

\renewcommand{\shortauthors}{F. Khoshnoud et al.}

\begin{abstract}
  \textbf{In pull-based development systems, code reviews and pull request comments play important roles in improving code quality. In such systems, reviewers attempt to carefully check a piece of code by different unit tests. Unfortunately, sometimes they miss bugs in their review of pull requests, which lead to quality degradations of the systems. In other words, disastrous consequences occur when bugs are observed after merging the pull requests. The lack of a concrete understanding of these bugs led us to investigate and categorize them. In this research, we try to identify missed bugs in pull requests of SmartSHARK dataset projects. Our contribution is twofold. First, we hypothesized merged pull requests that have code reviews, code review comments, or pull request comments after merging, may have missed bugs after the code review. We considered these merged pull requests as candidate pull requests having missed bugs. Based on our assumption, we obtained 3,261 candidate pull requests from 77 open-source GitHub projects. After two rounds of restrictive manual analysis, we found 187 bugs missed in 173 pull requests. In the first step, we found 224 buggy pull requests containing missed bugs after merging the pull requests. Secondly, we defined and finalized a taxonomy that is appropriate for the bugs that we found and then found the distribution of bug categories after analysing those pull requests all over again. The categories of missed bugs in pull requests and their distributions are: semantic (51.34\%), build (15.5\%), analysis checks (9.09\%), compatibility (7.49\%), concurrency (4.28\%), configuration (4.28\%), GUI (2.14\%), API (2.14\%), security (2.14\%), and memory (1.6\%).}
\end{abstract}



\maketitle

\section{Introduction}

Code reviews and pull request comments, as careful checks of source code by developers, are important parts of modern software development \cite{han2021understanding}. These techniques are also recognized as valuable tools for reducing software issues (e.g., bugs) in the code and improving the quality of software projects \cite{ackerman1984software, ackerman1989software}. Notably, in the GitHub community, developers can fork from an existing project into a new repository and then change parts of it, which are their interest. Using a pull request, they can submit the changes to that project. During pull request submission, the code reviewers are responsible for review of the code carefully before accepting and merging the changes as a contribution \cite{rahman2016correct}. 

Unfortunately, sometimes code reviewers miss bugs in their review of pull requests, which leads to quality degradations of the systems. In other words, disastrous consequences occur in the system, when bugs are found after merging a pull request. There is no still a concrete understating of these bugs which encouraged us to analyze and categorize them. Our research tries to formulate the following research questions to reach this goal: 

\begin{figure*}[ht]
    \centering
    \includegraphics[scale=0.83]{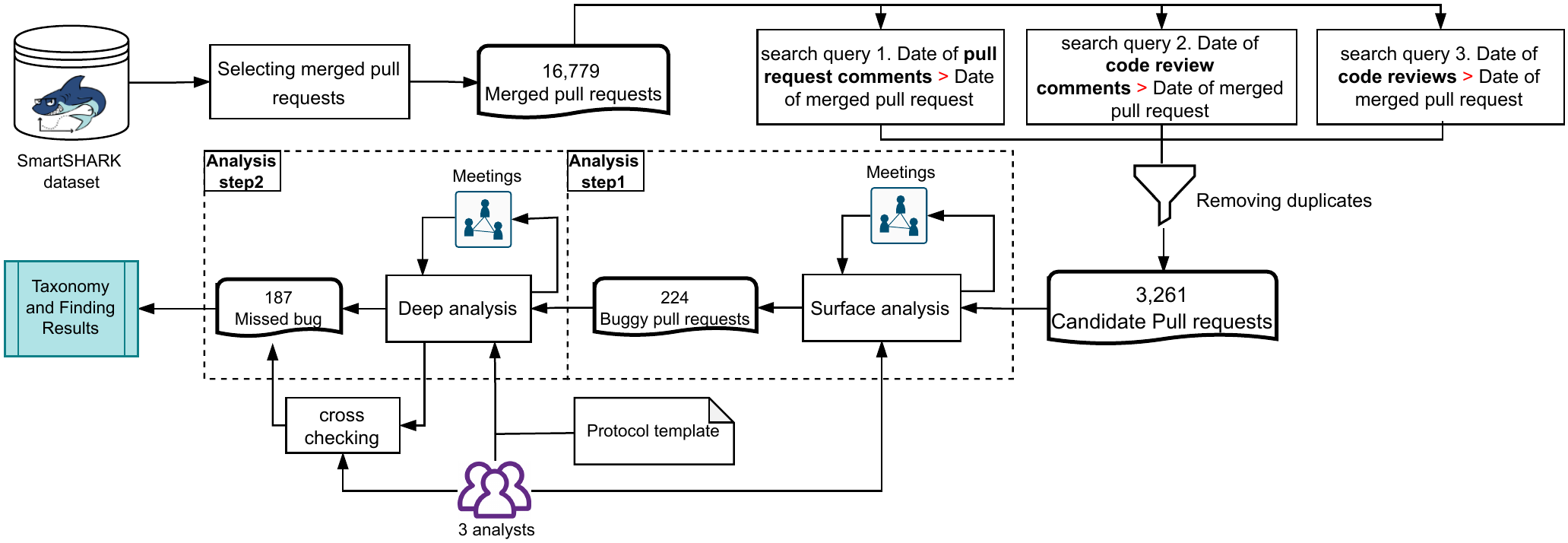}
    \caption{An overview of our study}
    \label{fig:overview}
\end{figure*}

\textbf{RQ1. \textit{How prevalent are possible missed bugs in code reviews?}} We gathered code reviews, code review comments, and pull request comments of 77 GitHub projects from the SmarkSHARK dataset after merging the pull requests. We gained 3,261 candidate pull requests. After analysing the candidate pull requests by three analysts, we identified 224 buggy pull requests that are missed in code reviews.


\textbf{RQ2. \textit{Which categories of bugs are missed in code reviews?}} The same analysts analyzed the buggy pull requests for the second time to find the types of bugs and then categorized them. This process led to 187 missed bugs from 173 pull requests which are categorized into ten common bug categories.

\textbf{RQ3. \textit{What are the distribution of the missed bugs in code reviews?}} After finalizing the taxonomy of missed bugs, we found the distributions of them. Our findings show semantic bug had 51.34\% as the highest distribution and memory bug had 1.6\% as the lowest distribution. Also, we publicly released the implementation, the protocol template, and the data collection used in this study available \cite{onlinedataset}.

\section{Related Work}

In this section, we discuss on relevant studies to this research.
Yu et al. \cite{yu2016reviewer} investigated the nature of Continuous Integration (CI) failures and social factors are associated with them and how they relate to eventual bugs. 
Zampetti et al. \cite{zampetti2019study_paper2} investigated interplay between pull request discussions and CI builds.
Tan et al. \cite{tan2014bug_paper7} studied the bug characteristics in pull requests of three large open-source software projects. In general, they found three types of bugs from the reports of code reviewers and contributors. 
WAN et al. \cite{wan2017bug_paper1} also studied the bug characteristics in eight open source blockchain systems hosted on GitHub.
Lastly, Han et al. \cite{han2021understanding} focused on understating code smells in code reviews by analysing 19,146 review comments. 

The closest works to our study are \cite{yu2016reviewer,zampetti2019study_paper2}. Although they worked on CI failures after merging the pull requests, we focus on a concrete understanding of types of bugs after merging the pull requests, which is still a challenge.

\section{Data Extraction and Collection}
Figure \ref{fig:overview} shows a complete overview of our study. The study uses SmarkSHARK dataset which is hosted on MongoDB database \cite{mongodb}. This dataset contains valuable information of 77 GitHub projects such as pull requests, code reviews, and etc. From this database, we first queried the merged pull requests of these projects. As shown in Figure \ref{fig:overview}, the query led to 16,779 merged pull requests. 

\begin{figure}[ht]
    \centering
    \includegraphics[scale=0.28]{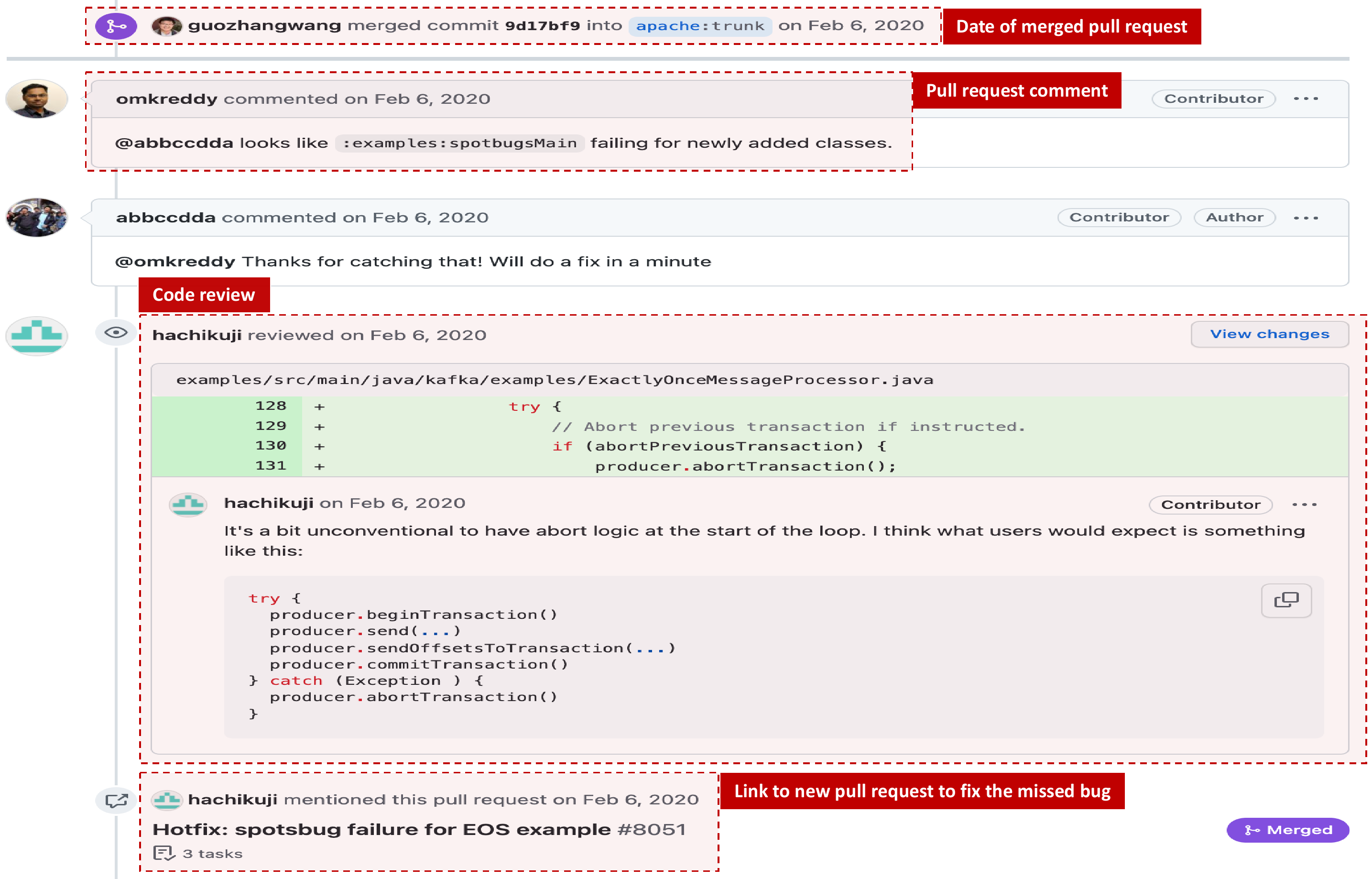}
    \caption{An example of missed bug in pull request $\#$8031 of kafka project}
    \label{fig:bugexample}
\end{figure}

\begin{figure*}[ht]
    \centering
    \includegraphics[scale=0.65]{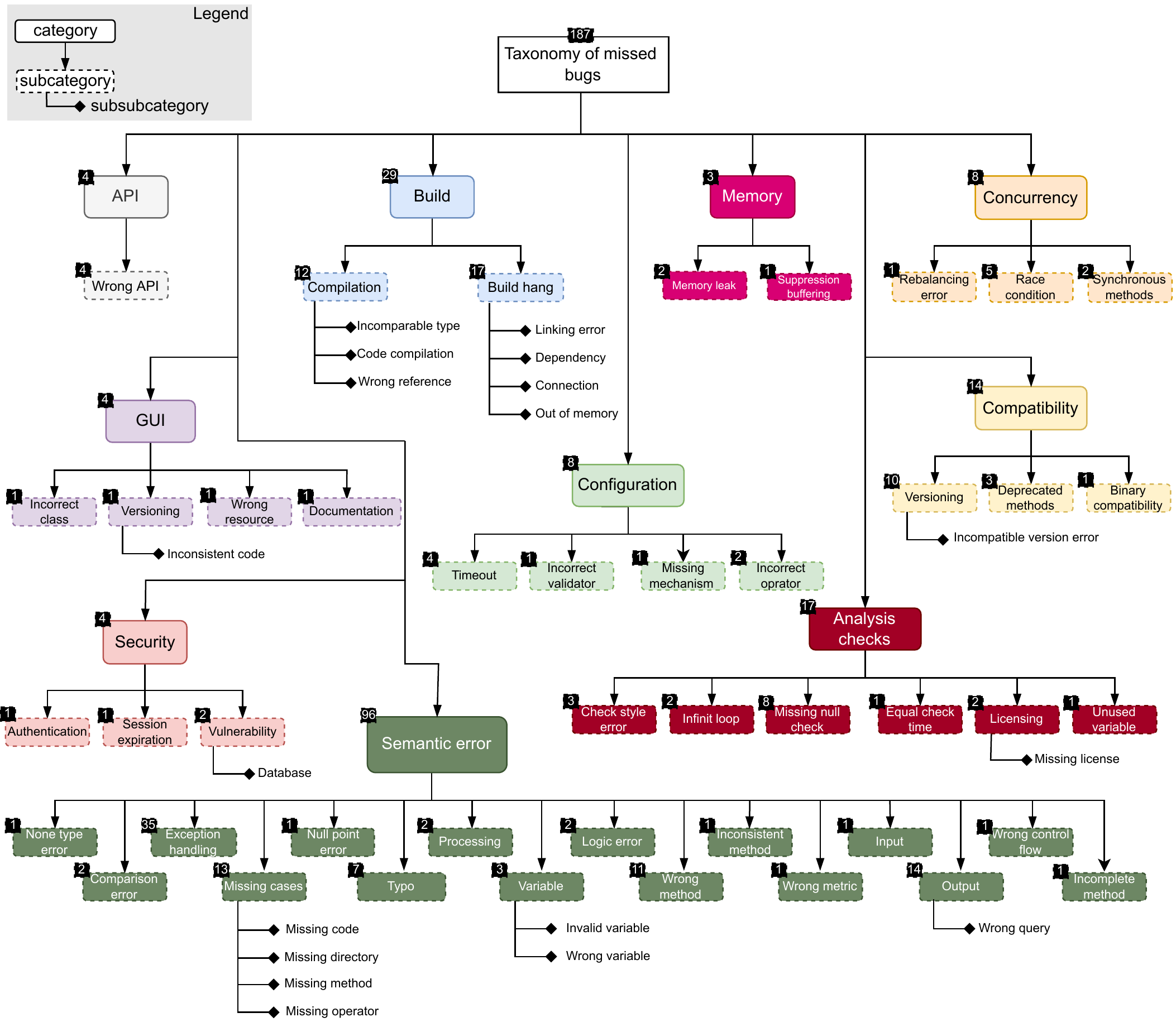}
    \caption{A taxonomy of missed bugs in code reviews}
    \label{fig:taxonomy}
\end{figure*}

We then ran three search queries which is described as follow: 

\textbf{Search query1.} Date of code reviews of a merged pull request must be more than the date of the merged pull request. This search shows 2,998 pull requests that have one or more code reviews after merging the pull requests.

\textbf{Search query2.} Date of code review comments of a merged pull request must be more than date of the merged pull requests. This results in 400 pull requests with one or more comments in code review comments. 

\textbf{Search query3.} Date of pull request comments must be more than date of a merged pull request. Result of this search shows 343 pull requests that have one or more pull request comments after merging the pull requests. 

After applying these search queries, we removed the duplicate pull requests. This resulted in 3,261 pull requests that may contain missed bugs after merging the pull requests. Note that we called these 3,261 pull requests as candidate pull requests.

\section{EXPERIMENTAL RESULTS}
\subsection{Methodology} \label{method}
Here we describe our manual analysis in two steps:

\textbf{Step1.} After extracting 3,261 candidate pull requests, three analysts analyzed these pull requests. Out of these 3,261 candidate pull requests, we assigned 1,087 candidate pull requests to each analyst to review and find buggy pull requests. They were asked to label each pull request that reports a bug and label those
that do not report any bugs. After reviewing 200 pull requests by each analyst, they held a meeting to decrease their ambiguities and dissimilarities. At the end of this analysis step, they attained 224 buggy pull requests which each buggy pull request had at least one bug report.

\textbf{Step2.} In this step, the same analysts were asked to deeply examine the buggy pull requests and find the type of bugs. Each analyst reviewed these 224 buggy pull requests independently. They first created a protocol template that is made of discovered bugs by \cite{tan2014bug_paper7, wan2017bug_paper1, zampetti2019study_paper2, humbatova2020taxonomy_paper3, islam2020repairing_paper4, jia2020empirical_paper5,  rahman2020exploratory_paper6} to find the best type for each bug reported in the buggy pull requests. The protocol template includes the bug categories and sub categories of seven manuscripts with their definitions. Second, they used this template to find and categorize the bugs reports of buggy pull requests.

After manual analysing, they also had several meetings in this step and they manually decided to mark some of the bugs reports of buggy pull requests as \textit{unknown} because of insufficient information. 51 out of 224 buggy pull requests led to \textit{unknown} label after reaching agreement among analysts. 
For the rest of buggy pull requests, we used Fleiss kappa \citep{fleiss1971measuring} to calculate the agreement between three analysts. The calculation of kappa value between the analysts was 0.78 meaning \textit{substantial agreement}.
Then, three analysts argued on their disagreements to reach consensus. The result was 187 bug reports from 173 buggy pull requests. Figure \ref{fig:bugexample} shows an example of missed bug after merging pull request $\#$8031 from kafka project. As shown in this Figure, the missed bug is fixed in pull request $\#$8051 as a hotfix pull request.

\begin{table*}[ht]
\caption{The category schema of types of missed bugs after merging the pull requests}\label{schema}
{\footnotesize
\begin{tabular}{|p{1.8cm}|p{13cm}|c|}
\hline
\textbf{Categories} & \textbf{Descriptions}    & \textbf{Distributions}                                                                                                                                                                                                                                                                                                                                                                                                       \\ \hline

Semantic          & This category of missed bugs caused by inconsistencies with the requirements or the programmers’ intention that do not belong to the other categories. For example, output corresponds to bugs when it displays wrong results \protect{\cite{wan2017bug_paper1, tan2014bug_paper7}}. & 51.34\% 

\\ \hline

Build             & This category of missed bugs caused by problems in the build, such as linking errors, connection errors, out of memory errors, or compilation errors. The problems may break the build \protect{\cite{wan2017bug_paper1, zampetti2019study_paper2}}. & 15.5\%

\\ \hline

Analysis checks   & This category of missed bugs caused by checks-related issues, licensing issues, or check style violations \protect{\cite{zampetti2019study_paper2}}.    & 9.09\%
                                    
\\ \hline

Compatibility     & This category corresponds to bugs when a system cannot normally run on a particular CPU architecture, operating system, or Web browser, etc. It also corresponds to the problems in the adaptation of the code (methods) to the new version of the library \protect{\cite{wan2017bug_paper1, islam2020repairing_paper4}}.   & 7.49\%                                                                                                 \\ \hline

Concurrency     & This category of missed bugs caused by synchronization problems in the system \protect{\cite{wan2017bug_paper1, jia2020empirical_paper5, tan2014bug_paper7}}.   & 4.28\%                                                                        
\\ \hline
                                            
Configuration   & This category of missed bugs caused by errors in dependent libraries, underlying operating systems, or non-code that affects functionality, including time configurations \protect{\cite{wan2017bug_paper1, zampetti2019study_paper2, jia2020empirical_paper5}}.     &   4.28\%                                                                                                                         \\ \hline

GUI               & This category of missed bugs caused by errors in graphical user interface, including incorrect class, incorrect information in documentation, inconsistent HTML code in different versions, displaying incorrect images, links, and color \protect{\cite{wan2017bug_paper1, rahman2020exploratory_paper6}}.   & 2.14\%                                                                                                          \\ \hline

API               & This category of missed bugs caused by problems arising from framework’s API usage such as wrong API usage \protect{\cite{humbatova2020taxonomy_paper3, islam2020repairing_paper4}}. & 2.14\%   

\\ \hline

Security         & This category of missed bugs caused by vulnerabilities of a part of the system, or bugs that violate confidentiality, integrity, or availability for the system \protect{\cite{wan2017bug_paper1, rahman2020exploratory_paper6, tan2014bug_paper7}}.      & 2.14\%                                                                                                                              
\\ \hline
                                                                        
Memory           & This category of missed bugs caused by improper handling of memory objects or memory usage problems \protect{\cite{wan2017bug_paper1, jia2020empirical_paper5, tan2014bug_paper7}}.   & 1.6\%                       
\\ \hline
\end{tabular}}
\end{table*}

\subsection{Results}
We identified 187 missed bugs from 173 buggy pull requests and then categorized them in a taxonomy of ten common bugs which are shown in Figure \ref{fig:taxonomy}. The missed bugs came from 28 open-source GitHub projects. Among these 28 projects containing the missed bugs, kafka, fineract, and calcite projects have the most missed bugs with 94, 20, and 17, respectively. Regarding the taxonomy, semantic bug has the most distribution (51.34\%) and memory bug has the least distribution (1.6\%). Figure \ref{fig:taxonomy} also provided the number of missed bugs in terms of category, subcategory, and subsubcategory. The category schema of missed bugs presented in Table \ref{schema} with their distributions. The following provides our categories of missed bugs with their examples. 

1) Semantic bug. \textbf{apache/kafka} \textbf{\href{https://github.com/apache/kafka/pull/8683}{$\#$8683}} reports the exception handling bug which is fixed in  \textbf{\href{https://github.com/apache/kafka/pull/8990}{$\#$8990}}. 

2) Build bug. \textbf{apache/kafka} \textbf{\href{https://github.com/apache/kafka/issues/8312}{$\#$8312}}  corresponds to removing a constructor caused to have broken the build. This fixed in \textbf{\href{https://github.com/apache/kafka/pull/8866}{$\#$8866}}. 

3) Analysis checks bug. \textbf{apache/kylin} \textbf{\href{https://github.com/apache/kylin/pull/1349}{$\#$1349}} reports that the apache license missed from a new file, which leads to fail the rat check. This missing license bug is fixed in \textbf{\href{https://github.com/apache/kylin/pull/1356}{$\#$1356}}. 

4) Compatibility bug. \textbf{apache/kafka} \textbf{\href{https://github.com/apache/kafka/pull/6188}{$\#$6188}} reports the incompatible version error and then this is fixed in pull request \textbf{\href{https://github.com/apache/kafka/pull/7101}{$\#$7101}}. 

5) Concurrency bug. \textbf{apache/kafka}
\textbf{\href{https://github.com/apache/kafka/pull/6178}{$\#$6178}} reports race condition bug and then \textbf{\href{https://github.com/apache/kafka/pull/6185}{$\#$6185}} fixes the bug with reverting that pull request. 

6) Configuration bug. \textbf{apache/kafka} \textbf{\href{https://github.com/apache/kafka/issues/6419}{$\#$6419}} reports timeout bug which occur in the session timeout. \textbf{\href{https://github.com/apache/kafka/issues/7072}{$\#$7072}} is a pull request that fixed it. 

7) GUI bug. We put \textbf{apache/kafka} \textbf{\href{https://github.com/apache/kafka/pull/7825}{$\#$7825}} in GUI bugs that reports using incorrect HTML class. The buggy pull request addressed in  pull request \textbf{\href{https://github.com/apache/kafka/pull/7837}{$\#$7837}}. 

8) API bug. \textbf{apache/commons-io} \textbf{\href{https://github.com/apache/commons-io/pull/124}{$\#$124}} had wrong API bug that is discussed in  \textbf{\href{https://issues.apache.org/jira/browse/IO-689}{$\#$jira}} as a bug report. The developers of commons-io project decided to fix it by updating javadocs and removing round trips. 

9) Security bug. \textbf{apache/tika} \textbf{\href{https://github.com/apache/tika/pull/219}{$\#$219}} uses jackson 2.9.4 which had a vulnerability. The vulnerability details in \href{https://nvd.nist.gov/vuln/detail/CVE-2018-7489}{CVE-2018-7489}. Thus, the developers of tika project opened \href{https://issues.apache.org/jira/browse/TIKA-2634}{TIKA-2634} issue in jira to fix the problem. 

10) Memory bug. \textbf{apache/kafka} \textbf{\href{https://github.com/apache/kafka/issues/8955}{$\#$8955}} reports memory leak bug that cause the usage of memory was increased. This is fixed in pull request \textbf{\href{https://github.com/apache/kafka/pull/9245}{$\#$9245}}.

\section{Discussion}

\textbf{Various categories of bugs are missed from code reviews.} 
Among 28 open-source GitHub projects, we found 187 bug reports after merging the pull requests which are categorized in ten different domain groups. Compared to Tan et al.'s work \cite{tan2014bug_paper7} and Wan et al.'s work \cite{wan2017bug_paper1}, their categories still exist in our categories of missed bugs, except for performance and hard Fork. \textbf{Semantic, build, and analysis checks bugs are dominant missed bugs in code reviews.} As shown in Figure \ref{fig:taxonomy}, semantic, build and analysis checks bugs  with the distribution of 51.34\%, 15.5\%, and 9.09\% are the most of reported bugs respectively that are missed after merging the pull requests. Exception handling bugs with 36.46\% out of semantic bugs, build hang bugs with 58.62\% out of build bugs, and missing null check with 47.06\% out of analysis checks bugs still significantly miss from code reviews after merging the pull requests. Overall our study recommends that code reviewers focus on reviewing the categories of missed bugs provided in this research (see Figure \ref{fig:taxonomy}) using relevant unit tests before merging the pull requests, especially the dominant missed bugs. Further research could be done to understand how is the reaction of code reviewers to the missed bugs in terms of time reaction or etc.

\section{Threats to validity}

\textbf{Construct Validity.} Two-step analysis of our study might have introduced concerns. In the first step, among 3,261 candidate pull requests, each analyst checked 1,087 candidate pull requests. We admit that they have missed some buggy pull requests having missed bugs. To mitigate this concern, they had a meeting after reviewing 200 candidate pull requests. In the second step, they categorized the missed bugs into different types of bugs. To mitigate this, they first used the protocol template (see section \ref{method}) and second they had several meetings on their disagreements to reduce this threat.

\textbf{External Validity.} Generalizability in this work might be a threat to external validity. SmartSHARK dataset only has projects from the apache domain. Although this domain is one of the popular domains, we acknowledge that our results may change if we use different domains such as Microsoft\footnote{\href{https://github.com/microsoft}{https://github.com/microsoft}}.

\section{CONCLUSION and FUTURE WORK}
Since the lack of understanding of missed bugs by code reviewers that may have hazardous consequences on other parts of a system. Hence, in this work, we found pull requests of projects from SmarkSHARK dataset which reports bugs after merging that pull requests. To this end, we first extracted 3,261 candidate pull requests which may report bugs in code reviews, code review comments, and pull request comments after merging. After analysing the candidate pull requests, we obtained 187 bug reports from 173 buggy pull requests and categorized them in ten types. At the end, we obtained a taxonomy containing categories, subcategories, and subsubcategories of missed bugs. The distribution of each category is provided. The final result shows us that the most missed bug is semantic (51.34\%) and the least missed bug is memory (1.6\%).
In future work, we plan to investigate different domains of projects like Microsoft domain to enrich our findings. Also, we would like to further analyze the missed bugs in different directions such as reaction time of reviewers.




\bibliographystyle{bib_style}
\bibliography{bib_file}


\end{document}